\documentclass[a4paper,11pt]{article}
\usepackage{pos}

\usepackage{subcaption}
\usepackage{float}
\restylefloat{table} % gives H position that is never moved
\usepackage[edges]{forest}
\usepackage{hyperref}
\usepackage{mathtools}

\title{Novel Algorithms for Computing Correlation Functions of Nuclei}
\author*[a]{Nabil Humphrey}

\author[b,c]{William Detmold}

\author[a]{Ross D. Young}

\author[a]{James M. Zanotti}

\affiliation[a]{CSSM, Department of Physics, University of Adelaide, \\
Adelaide SA 5005, Australia}

\affiliation[b]{Center for Theoretical Physics, \\
Massachusetts Institute of Technology, \\
Cambridge, MA 02139, USA}

\affiliation[c]{The NSF AI Institute for Artificial Intelligence and Fundamental Interactions}

\emailAdd{nabil.humphrey@adelaide.edu.au}

\dedicated{%
    \vspace{8 mm}
    MIT-CTP/5381\\
    ADP-21-24/T1171
}

\abstract{The computational cost required to calculate nuclear correlation functions grows factorially in the number of quarks, making the study of large nuclei inaccessible to \textit{ab initio} study using lattice QCD at the present time. However, the tensor expressions corresponding to many of these correlation functions exhibit a high degree of permutation symmetry that can be exploited to reduce computational work. We present promising speed-ups for certain choices of interpolating operators using two new algorithms for computing nuclear correlation functions.}

\FullConference{%
 The 38th International Symposium on Lattice Field Theory, LATTICE2021
  26th-30th July, 2021
  Zoom/Gather@Massachusetts Institute of Technology
}

\begin{document}

\maketitle

\section{Introduction}
This work will explore the use of lattice QCD in the \textit{ab initio} study of the properties and interactions of multi-hadron systems directly from QCD. There are deep physical and numerical challenges in this regard, including a proliferation of possibilities for interpolating operators that couple to low energy states, factorial scaling of contraction complexity in general, signal-to-noise scaling \cite{parisi_1983, lepage_1989, beane_2009}, and potentially dominating floating-point precision errors \cite{detmold_2008}. The focus of this work is the computational resource scaling associated with numerically evaluating correlation functions built from interpolating operators possessing the quantum numbers of a multi-hadron system. For a given set of quantum numbers, a two-point correlation function takes the form:
\begin{align}
    C &= \langle \mathscr{O}' \mathscr{O}^\dagger \rangle =  \sum_{(\vec a, \vec a') \in \mathcal{I}} W^{a_1, \dots, a_{n_q}}_{a_1', \dots, a_{n_q}'}\   \left\langle q(a_1) \dots q(a_{n_q}) \overline{q}(a_1') \dots \overline{q}(a_{n_q}') \right \rangle ,
    \label{eqn:correlator}
\end{align}
\noindent where $\vec a = (a_1, \dots, a_{n_q})$ combine flavour $f_i$, colour indices $c_i$, spinor indices $\alpha_i$, and spacetime indices $x_i$ of quark fields $q(a_i) \equiv q^{f_i, c_i}_{\alpha_i}(x_i)$; $W^{a_1, \dots, a_{n_q}}_{a_1', \dots, a_{n_q}'}$ combines tensors in $\mathscr{O}$ which project onto the correct set of quantum numbers; and $n_q = n_u + n_d$, where $n_u$ ($n_d$) denotes the number of up (down) quark fields in operator $\mathscr{O}$. The set of values over which $a_i, a_j'$ range, given by $\mathcal{I} = \{ (\vec a, \vec a') \mid W^{\vec a}_{\vec a'} \neq 0 \}$, is referred to as the \textit{index set} and its cardinality $\left| \mathcal{I} \right|$ is referred to as the \textit{index size}. The number of floating-point operations required to directly evaluate $C$ scales as the product of the index size $\left| \mathcal{I} \right|$ and the number of quark-level Wick contractions, which is given by $n_u! n_d!$ in the case where $\mathscr{O}$ contains only quark fields rather than their adjoints. \par
In order to construct nuclear correlation functions, we must first construct quark-level interpolating operators with the desired set of quantum numbers. A convenient approach for multi-baryon systems is to combine local baryon interpolating operators by tying together uncontracted spinor indices. We consider the following set of single baryon operators (examples of $\mathscr{O}$):
\begin{align*}
    p^\alpha(x) &= \epsilon_{abc} (u_a^T(x) (C\gamma_5) d_b(x)) u_c^\alpha(x) \qquad \text{(proton)} \\
    n^\alpha(x) &= \epsilon_{abc} (d_a^T(x) (C\gamma_5) u_b(x)) d_c^\alpha(x) \qquad \text{(neutron)} \\
    p^\alpha_\pm(x) &= \epsilon_{abc} (u_a^T(x) (C\gamma_5 P_\pm) d_b(x)) u_c^\alpha(x) \qquad \text{(non-relativistic proton)} \\
    n^\alpha_\pm(x) &= \epsilon_{abc} (d_a^T(x) (C\gamma_5 P_\pm) u_b(x)) d_c^\alpha(x) \qquad \text{(non-relativistic neutron),}
\end{align*}
where $P_{\pm} = \frac{1}{2} (1 \pm \gamma_4)$. From single baryon operators, we may construct a selection of multi-baryon operators (examples of $\mathscr{O}$):
% \begin{align}
%     &\text{Dinucleon I:} D_I(x) = n^T(x) (C\gamma_5) p(x) \notag \\
%     &\text{Dinucleon II:} D_{II}(x) = \frac{1}{\sqrt{2}} \left[ n^T(x) (C\gamma_5) p(x) - p^T(x) (C\gamma_5) n(x) \right] \notag\\
%     &\text{Helium-3 I:} {}^3 He^j_I(x) = p_-^T(x) (C\gamma_5) n_+(x) p_+^j(x) \label{eqn:ops}\\
%     &\text{Helium-3 II:} {}^3 He^j_{II}(x) = \frac{1}{\sqrt{6}} \left[ p_-^T(x) (C\gamma_5) n_+(x) p_+^j(x) - p_+^T(x) (C\gamma_5) n_+(x) p_-^j(x) \right.\notag \\
%     &\qquad \qquad \qquad \qquad \qquad \qquad \left. + n_+^T(x) (C\gamma_5) p_+(x) p_-^j(x) - n_+^T(x) (C\gamma_5) p_-(x) p_+^j(x) \right. \notag\\
%     &\qquad \qquad \qquad \qquad \qquad \qquad \left. + p_+^T(x) (C\gamma_5) p_-(x) n_+^j(x) - p_-^T(x) (C\gamma_5) p_+(x) n_+^j(x) \right] \notag
% \end{align}
\begin{align}
    &D_I(x) = n^T(x) (C\gamma_5) p(x) \qquad \qquad \qquad \qquad \qquad \qquad \quad \text{(Dinucleon I)} \notag \\
    &D_{II}(x) = \frac{1}{\sqrt{2}} \left[ n^T(x) (C\gamma_5) p(x) - p^T(x) (C\gamma_5) n(x) \right] \qquad \text{(Dinucleon II)} \notag\\
    &{}^3 He^j_I(x) = p_-^T(x) (C\gamma_5) n_+(x) p_+^j(x) \qquad \qquad \qquad \qquad \quad \ \ \text{(Helium-3 I)} \label{eqn:ops}\\
    &{}^3 He^j_{II}(x) = \frac{1}{\sqrt{6}} \left[ p_-^T(x) (C\gamma_5) n_+(x) p_+^j(x) - p_+^T(x) (C\gamma_5) n_+(x) p_-^j(x) \right.\notag \\
    &\qquad \qquad \qquad \quad \left. + n_+^T(x) (C\gamma_5) p_+(x) p_-^j(x) - n_+^T(x) (C\gamma_5) p_-(x) p_+^j(x) \right. \notag\\
    &\qquad \qquad \qquad \quad \left. + p_+^T(x) (C\gamma_5) p_-(x) n_+^j(x) - p_-^T(x) (C\gamma_5) p_+(x) n_+^j(x) \right] \qquad \text{(Helium-3 II)} \notag \\
    &\text{(note that $D_{II}$ and ${}^3He_{II}$ project spin and isospin quantum numbers while $D_I$ and ${}^3He_I$ do not).} \notag
\end{align}
\indent The literature explores and evaluates a number of existing algorithms to reduce the numerical work associated with multi-hadron contractions. The formulation into \textit{hadron blocks}, as described and evaluated in \autoref{sec:HadronBlocks}, has been explored in Refs.\ \cite{detmold_2013,doi_2013,gunther_2013,numera_2017} in the context of nuclei. The construction of \textit{Index Lists}, as described in Ref.\ \cite{doi_2013}, pre-computes the minimal subset of the index set of a numerically expensive tensor which has a non-vanishing contribution to the correlator. We construct and use Index Lists in our implementation of hadron blocks in \autoref{sec:HadronBlocks}. The matrix determinant formulation, as devised in Ref.\ \cite{detmold_2013}, re-casts the quark-level Wick contraction $\left\langle q(a_1) \dots q(a_{n_q}) \overline{q}(a_1') \dots \overline{q}(a_{n_q}') \right\rangle$ as the determinant of a block matrix containing quark propagator elements $S(a_i; a_j') \equiv \underbracket{q(a_i) \overline{q}}(a_j')$. When combined with LU factorisation \cite{lindfield_2019_73}, the asymptotic scaling of the number of operations is reduced from $\mathcal{O}\left(n_u! n_d! \right)$ to $\mathcal{O}\left(n_u^3 n_d^3 \right)$. The index size scaling, which for light nuclei is the dominant scaling factor, is left unchanged.

\section{Hadron Blocks} \label{sec:HadronBlocks}

\begin{figure}
    % \begin{subfigure}{0.5\textwidth}
    %     \centering
    %     \includegraphics[width=\textwidth]{images/TalkPlot_Deuteron_NaiveVsBlock_LargeVol.png}
    % \end{subfigure}%
    % \begin{subfigure}{0.5\textwidth}
    %     \centering
    %     \includegraphics[width=\textwidth]{images/TalkPlot_Helium4_NaiveVsBlock_LargeVol.png}
    % \end{subfigure}
    \centering
    \includegraphics[width=\textwidth]{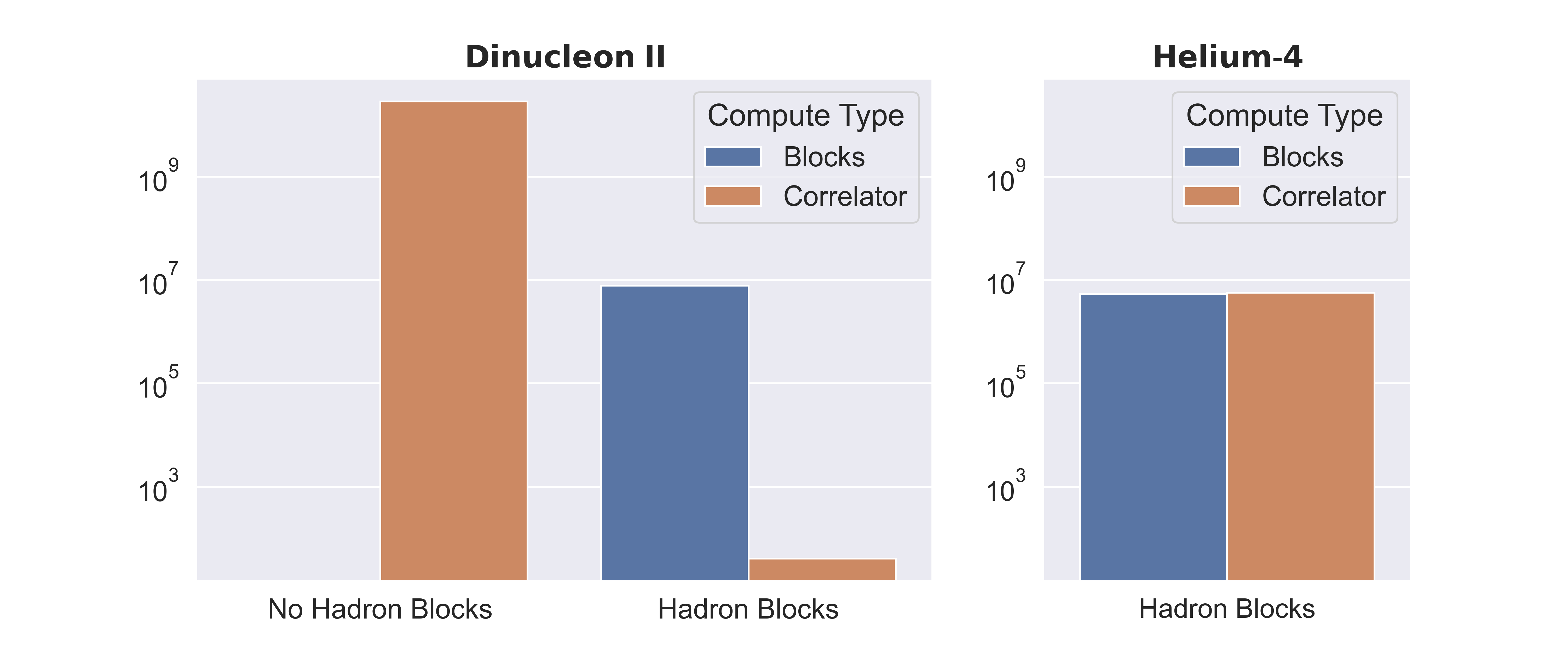}
    \caption{Hadron block benchmark, measuring wall-clock time in milliseconds on a single core of an Intel Xeon Scalable Cascade Lake processor; lattice volume $64^3$. The helium-4 correlator without hadron blocks was too costly to perform here, but an upper bound on the number of floating point operations required is $10^{22}$.}
    \label{fig:hadron_blocks}
\end{figure}

The formulation of multi-hadron correlation function contractions in terms of single hadron building blocks has been explored in Refs.\ \cite{detmold_2013,doi_2013,gunther_2013,numera_2017}. A hadron block is a tensor, $f^h_{\vec p}(x', \vec \xi)$, constructed from the constituent quarks of a hadron $h$ with colour and spinor indices bundled into $\vec \xi$, created at a fixed source spacetime location $x'$ and annihilated as a (momentum-projected to $\vec p$ and colour-contracted) hadron at the sink. In the case of a single proton block, $f^P_{\vec p}(x', \vec \xi)$, we can compute the necessary Wick contractions using explicit colour and spinor indices for both the hadron block $\left(\vec \xi = (a', \beta', b', \gamma', c', \alpha', \alpha) \right)$ and the quark propagator $\left(S^{f,cc'}_{\alpha \alpha'}(x,x') \equiv \underbracket{q^{f, c}_\alpha(x) \overline{q}}{}^{f, c'}_{\alpha'}(x') \right)$ via:
\begin{align}
    &f^P_{\vec p}(x', \vec \xi) \equiv f^P_{\vec p}(x',a',\beta',b',\gamma',c',\alpha')_\alpha \nonumber \\
    &\qquad \coloneqq \left\langle \sum_{x} e^{-i \vec p \cdot \vec x} p_{\alpha}(x) \overline{u}^{a'}_{\beta'}(x') \overline{d}^{b'}_{\gamma'}(x') \overline{u}^{c'}_{\alpha'}(x') \right\rangle \nonumber\\
    &\qquad = \sum_{x} e^{-i \vec p \cdot \vec x} \epsilon^{abc} \Gamma_{\beta \gamma} \left\langle u^a_\beta(x) d^b_\gamma(x) u^c_\alpha(x) \overline{u}^{a'}_{\beta'}(x') \overline{d}^{b'}_{\gamma'}(x') \overline{u}^{c'}_{\alpha'}(x') \right\rangle \nonumber\\
    &\qquad = \sum_{x} e^{-i \vec p \cdot \vec x} \epsilon^{abc} \Gamma_{\beta \gamma} \left[ S^{u,a c'}_{\beta \alpha'}(x,x') S^{u,c a'}_{\alpha \beta'}(x,x') - S^{u,a a'}_{\beta \beta'}(x,x') S^{u,c c'}_{\alpha \alpha'}(x,x') \right] S^{d,b b'}_{\gamma \gamma'}(x,x').
\end{align}
\indent The primary advantage of this construction is that block expressions are typically re-used many times during the course of evaluating particular choices of multi-hadron correlators. The factorial number of Wick contractions in the correlator is suppressed by a factor of $2^A$ for $A$ baryons, but without block expression re-use this computation cost is merely transferred to the hadron block evaluation. To measure the performance improvement associated with using hadron blocks, Figure \ref{fig:hadron_blocks} compares wall-clock time for the correlator computation (excluding propagator computation) on a $64^3$ volume using the dinucleon II operator in Eq. (\ref{eqn:ops}) and the helium-4 operator given in Ref.\ \cite{yamazaki_2010}.

Figure \ref{fig:hadron_blocks} demonstrates that hadron blocks offer a clear performance improvement, even for two-hadron systems. Note that the cost to compute the block expressions (shown in blue) remains constant\footnote{Hadron Block cost differs between relativistic (e.g. dinucleon II) and non-relativistic (e.g. helium-4) forms.} in baryon number, whilst the cost to contract these blocks to form the multi-hadron correlator (shown in brown) scales factorially. The block expression cost does not benefit significantly from further symmetry-based optimisation, so the focus of the remaining sections will be the correlator cost itself (shown in brown).

\section{Empirical Symmetry Properties} \label{sec:SymProperties}

\begin{figure}
    \centering
    % \begin{subfigure}{0.5\textwidth}
    %     \centering
    %     \includegraphics[width=\textwidth]{images/TalkPlot_Saturation.png}
    % \end{subfigure}%
    % \begin{subfigure}{0.5\textwidth}
    %     \centering
    %     \includegraphics[width=\textwidth]{images/TalkPlot_Saturation_Helium3.png}
    % \end{subfigure}
    \includegraphics[width=0.7\textwidth]{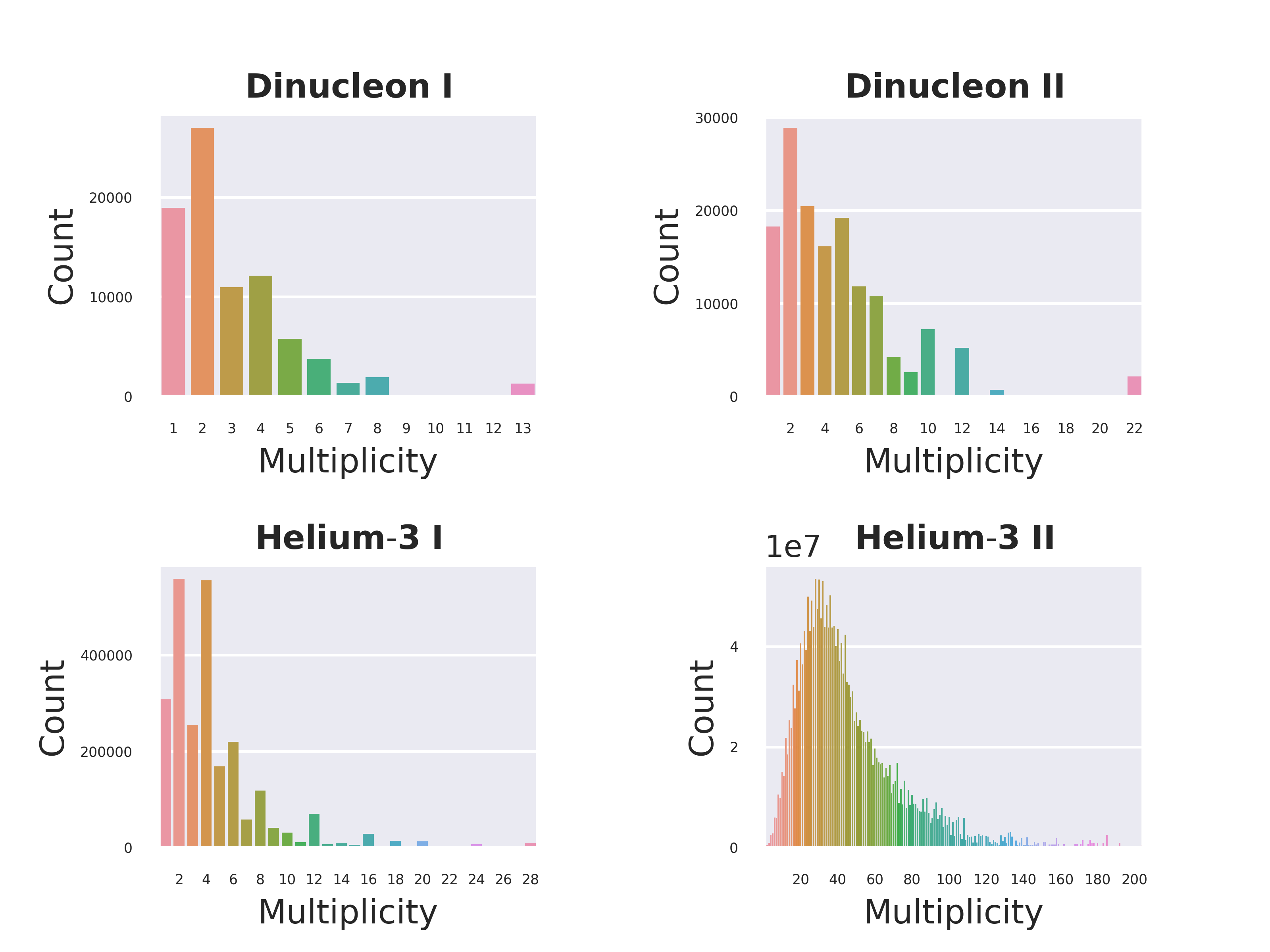}
    \caption{Multiplicity histograms for dinucleon I/II and helium-3 I/II operators in Eq. (\ref{eqn:ops}), where multiplicities are computed as the number of identical terms in the expansion of Eq. (\ref{eqn:gen_form}) after canonical ordering of terms.}
    \label{fig:multiplicity_histograms}
\end{figure}

The following investigates some of the term-wise symmetry properties of the particular multi-baryon correlators under consideration. Under the hadron block construction, correlators of $A$ baryons take the general form:
\begin{align}
    C &= \sum_{k=1}^{N_w} w_k \epsilon_{a_1 b_1 c_1} \dots \epsilon_{a_{A} b_{A} c_{A}} \Gamma_{\alpha_1 \beta_1}^{(1,k)} \dots \Gamma_{\alpha_{N_\Gamma} \beta_{N_\Gamma}}^{(N_\Gamma,k)} f^{h_1}_{\vec p_{(1,k)}} \left( x_{(1,k)}^{'}, \vec \xi_{(1,k)} \right) \dots f^{h_{A}}_{\vec p_{(A,k)}} \left( x_{(A,k)}^{'}, \vec \xi_{(A,k)} \right), \label{eqn:gen_form}
\end{align}
where $w_k \in \mathbb{C}$, $N_\Gamma$ is the number of gamma matrices subject to $A \leq N_\Gamma \leq 2 A$, $N_w$ is the number of terms in the contracted expression for $C$, $\left\{ f^h_{\vec p}(x', \vec \xi) \right\}$ are the sink-momentum projected hadron block functions as constructed in \autoref{sec:HadronBlocks}, $\{ h_i \}$ are hadrons determined from $\mathscr{O}$, $\left\{ x_{i,k}^{'} \right\}$ are the (possibly different) source spacetime points for the $k^{th}$ term in the correlator, and $\left\{ \vec \xi_{(i,k)} \right\}$ are functions of $\{ a_i \}$, $\{ b_i \}$, $\{ c_i \}$, $\{ \alpha_i \}$, $\{ \beta_i \}$ so that $\left\{ f^h_{\vec p}(x', \vec \xi) \right\}$ are fully contracted with $\{ \epsilon_{abc} \} \cup \{ \Gamma_{\alpha \beta} \}$. For example, the correlator for dinucleon I in Eq. (\ref{eqn:ops}) has $\mathscr{O} = D_I(x)$, $A = 2$, $N_\Gamma = 4$, $N_w = 9$, $h_1 = P$, $h_2 = N$, and the $k=1$ term for Eqn. (\ref{eqn:gen_form}) is:
\begin{align}
    &\epsilon_{a_1 b_1, c_1} \epsilon_{a_2 b_2 c_2} (C\gamma_5)_{\alpha_1 \beta_1} (C\gamma_5)_{\alpha_2 \beta_2} (C\gamma_5)_{\alpha_3 \beta_3} (C\gamma_5)_{\alpha_4 \beta_4} \times \nonumber \\
    &\qquad \qquad \qquad \qquad \qquad \times f^P_{\vec 0}(x', a_1, \alpha_2, b_1, \beta_2, c_1, \alpha_3)_{\alpha_1} f^N_{\vec 0}(x', a_2, \alpha_4, b_2, \beta_4, c_2, \beta_3)_{\beta_1}.
\end{align}
When all internal indices $\{ a_i \}$, $\{ b_i \}$, $\{ c_i \}$, $\{ \alpha_i \}$, $\{ \beta_i \}$ are summed over, we are left with strings of $f^h_{\vec q}(\widetilde{x}\ {}^{'}, \vec \eta)$ factors: 
\begin{align}
    C &= \sum_{k=1}^{\widetilde{N}_w} \widetilde{w}_k f^{h_1}_{\vec q_{(1,k)}} \left( \widetilde{x}_{(1,k)}^{\ '}, \vec \eta_{(1,k)} \right) \dots f^{h_{A}}_{\vec q_{(A,k)}} \left( \widetilde{x}_{(A,k)}^{\ '}, \vec \eta_{(A,k)} \right), \label{eqn:expanded_form}
\end{align}
where $\vec \eta_{i,k}$ are bundles of fixed colour/spinor values (note the notation change $\vec \xi \to \vec \eta$ to differentiate contracted and fixed index bundles),  $\widetilde{w}_k$ are the new weights after expansion, $\widetilde{N}_w$ is the number of terms in the expanded expression for $C$, and $\left\{ \widetilde{x}_{i,k}^{\ '} \right\}$ are the source spacetime points of the expanded expression (noting that $\widetilde{x}_{i,k}^{\ '} \neq x_{i,k}^{\ '}$ in general). For example, the correlator for dinucleon I in Eq. (\ref{eqn:ops}) has the $k = 1$ term for Eqn. (\ref{eqn:expanded_form}) given by:
\begin{align}
    f^P_{\vec 0}(x', 1, 1, 2, 2, 3, 1)_1 f^N_{\vec 0}(x', 1, 1, 2, 2, 3, 2)_2.
\end{align}
We may then canonically order the factors $f^h_{\vec q}(\widetilde{x}\ {}^{'}, \vec \eta)$ using lexicographic order on $(h, \vec q, \widetilde{x}\ {}^{'}, \vec \eta)$ tuples, and assign multiplicities to identical terms. Performing this process with the operators for dinucleon I/II and helium-3 I/II in Eq. (\ref{eqn:ops}) yields multiplicity histograms in Figure \ref{fig:multiplicity_histograms}. The multiplicity power law relationship seen in Figure \ref{fig:multiplicity_histograms} suggests that significant computational savings can be made by computing each degenerate term once, adjusting the coefficients by multiplicity. This idea also motivates \textit{factor trees} as explored in \autoref{sec:FactorTrees}.

\section{Factor Trees} \label{sec:FactorTrees}

Starting with the observations of empirical symmetry properties, as explored in \autoref{sec:SymProperties}, together with the property that the set of factors $\{ f^h_{\vec q}(\widetilde{x}\ {}^{'}, \vec \eta) \}$ is small compared with the set of terms in $C$, we can expect a high degree of both term-wise and factor-wise `compression' in multiplicity accumulation and factorisation. To target expressions of this type we construct \textit{factor trees}, in which the sum of all root-to-leaf paths through the tree, weighted by the leaf coefficients, is equal to the original tensor expression. 
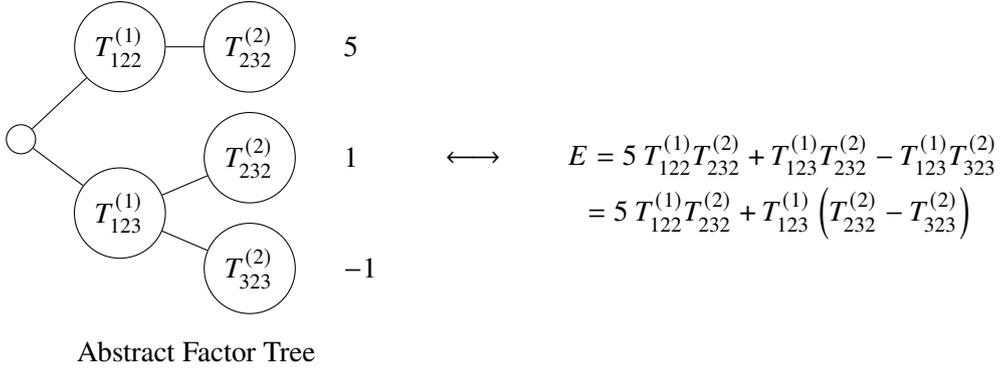
\begin{figure}
    \centering
    \begin{align*}
        \begin{gathered}
        \begin{forest}
        for tree={minimum size=1em,grow'=0}
        [,circle,draw
            [$T^{(1)}_{122}$,circle,draw
                [$T^{(2)}_{232}$,circle,draw [$5$,no edge]]
            ]
            [$T^{(1)}_{123}$,circle,draw
                [$T^{(2)}_{232}$,circle,draw [$1$,no edge]],
                [$T^{(2)}_{323}$,circle,draw [$-1$,no edge]]
            ]
        ]
        \end{forest}\\
        \text{Abstract Factor Tree}
        \end{gathered}
        \begin{gathered}
        \qquad \longleftrightarrow \qquad E = 5\ T^{(1)}_{122} T^{(2)}_{232} + T^{(1)}_{123} T_{232}^{(2)} - T^{(1)}_{123} T_{323}^{(2)} \\
        \qquad \qquad \qquad = 5\ T^{(1)}_{122} T^{(2)}_{232} + T^{(1)}_{123} \left( T_{232}^{(2)} - T_{323}^{(2)} \right) 
        \end{gathered}
    \end{align*}
    \caption{Abstract Factor Tree example (left) with corresponding tensor expression $E$ (right). $T_{ijk}^{(1)}$, $T_{ijk}^{(2)}$ are rank-3 tensors.}
    \label{fig:abstract_factor_tree}
\end{figure}

\begin{figure}
    \centering
    \begin{align*}
        \begin{gathered}
        \begin{forest}
        for tree={minimum size=1em,grow'=0}
        [,circle,draw
            [$T^{(1)}_{122}$,circle,draw
                [$T^{(2)}_{232}$,circle,draw [$5$,no edge]]
            ]
            [$T^{(1)}_{123}$,circle,draw
                [$T^{(2)}_{232}$,circle,draw [$1$,no edge]],
                [$T^{(2)}_{323}$,circle,draw [$-1$,no edge]]
            ]
        ]
        \end{forest}
        \end{gathered}
        \qquad &\longleftrightarrow \qquad 
        \begin{gathered}
        \textit{factors} = \left[ \circ, T^{(1)}_{122}, T^{(2)}_{232}, T^{(1)}_{123}, T^{(2)}_{232}, T^{(2)}_{323} \right] \\
        \textit{children} = [2, 1, 0, 2, 0, 0] \\
        \textit{coefficients} = [5, 1, -1] 
        \end{gathered}
    \end{align*}
    \begin{align*}
    \begin{gathered}
        \text{(a) Abstract Factor Tree}
    \end{gathered}
    \qquad \qquad \qquad \qquad \qquad \quad 
    \begin{gathered}
        \text{(b) Linearised Factor Tree}
    \end{gathered}
    \end{align*}
    \caption{Relationship between an abstract factor tree (a) and the corresponding linearised factor tree (b). Example as in Figure \ref{fig:abstract_factor_tree}.}
    \label{fig:lin_factor_tree}
\end{figure}
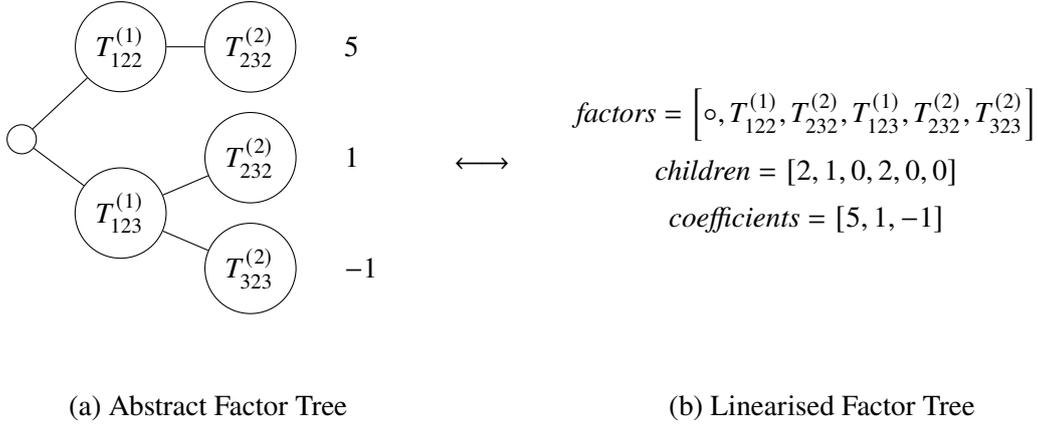

A natural way to represent an abstract factor tree that corresponds to a tensor expression $E$ as in Figure \ref{fig:abstract_factor_tree} on a computer might be to represent each node as a structure and to form links between the nodes using pointers. This would be highly inefficient in two senses: first, it doesn't exploit the predictable access pattern (depth-first traversal) that evaluation of the factor trees entails and would as a result be significantly memory access bound; and second, it would have a large memory footprint, taken up almost entirely by pointers of $parent \to child$ relationships. A representation that is much more efficient on both counts is given by a \textit{linearised factor tree}. As shown in Figure \ref{fig:lin_factor_tree}, a linearised factor tree is represented by three arrays: \textit{factors}, \textit{children}, and \textit{coefficients}. The \textit{factors} array is given by the depth-first traversal of the vertices of the abstract factor tree. The \textit{children} array is given by the number of children of the vertices in depth-first traversal order. The \textit{coefficients} array is the list of leaf coefficients in depth-first traversal order. 

% \begin{figure}
%     \centering
%     \begin{align*}
%         \begin{gathered}
%         \begin{forest}
%         for tree={minimum size=1em,grow'=0}
%         [,circle,draw
%             [$T^{(1)}_{122}$,circle,draw
%                 [$T^{(2)}_{232}$,circle,draw [$5$,no edge]]
%             ]
%             [$T^{(1)}_{123}$,circle,draw
%                 [$T^{(2)}_{232}$,circle,draw [$1$,no edge]],
%                 [$T^{(2)}_{323}$,circle,draw [$-1$,no edge]]
%             ]
%         ]
%         \end{forest}\\
%         \text{(a) Abstract Factor Tree}
%         \end{gathered}
%         \qquad &\longleftrightarrow \qquad 
%         \begin{gathered}
%         \textit{factors} = \left[ \circ, T^{(1)}_{122}, T^{(2)}_{232}, T^{(1)}_{123}, T^{(2)}_{232}, T^{(2)}_{323} \right] \\
%         \textit{children} = [2, 1, 0, 2, 0, 0] \\
%         \textit{coefficients} = [5, 1, -1] \\ \\ \\ \\
%         \text{(b) Linearised Factor Tree}
%         \end{gathered}
%     \end{align*}
%     \caption{Relationship between an abstract factor tree (a) and the corresponding linearised factor tree (b). Example as in Figure \ref{fig:abstract_factor_tree}.}
%     \label{fig:lin_factor_tree}
% \end{figure}

\begin{figure}
    \centering
    \includegraphics[width=0.9\textwidth]{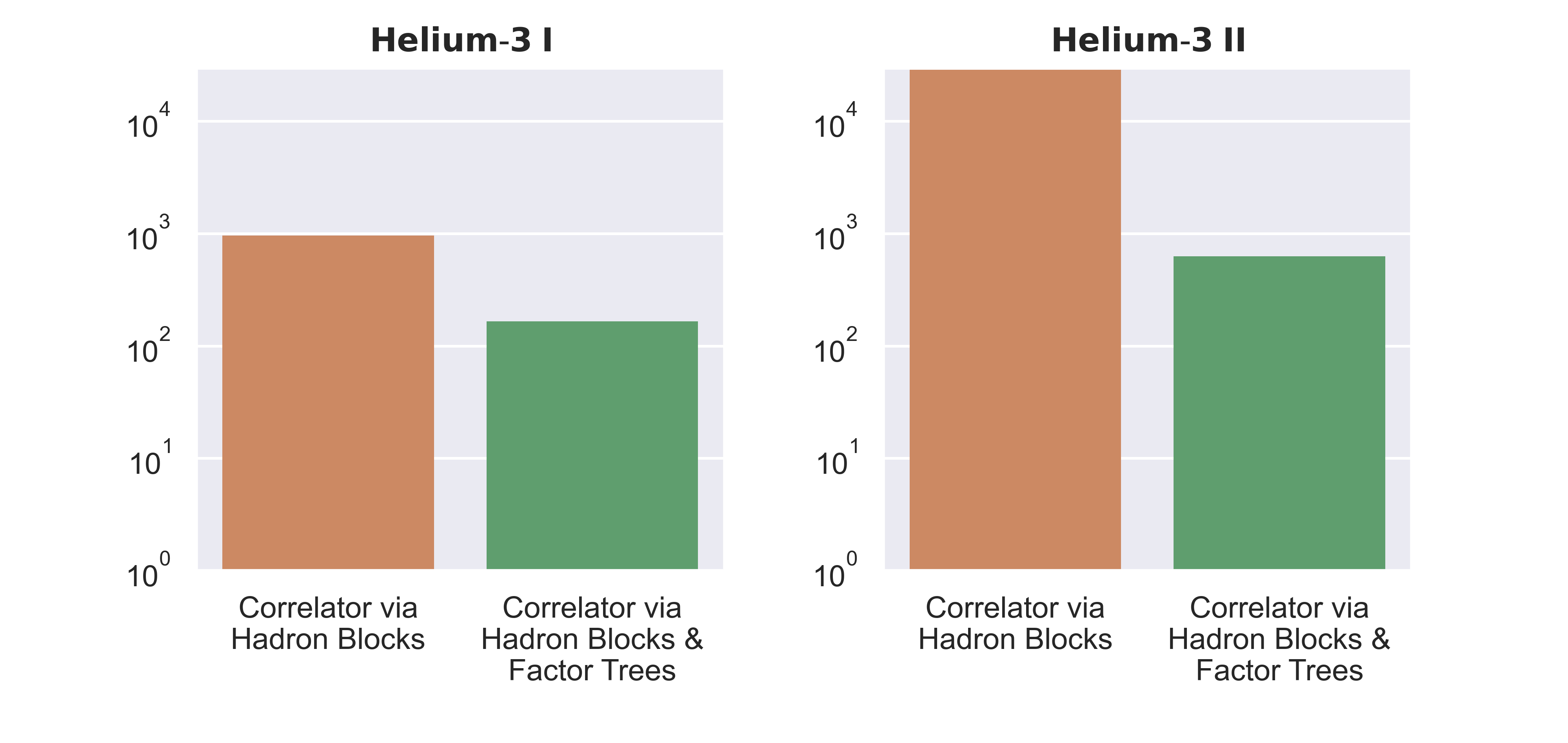}
    \caption{Factor tree benchmark, measuring wall-clock time in milliseconds on a single core of an Intel Xeon Scalable Cascade Lake processor; lattice volume $64^3$.}
    \label{fig:tree_benchmark}
\end{figure}

To measure the performance improvement associated with using factor trees, Figure \ref{fig:tree_benchmark} compares wall-clock time for correlator computation (excluding propagator and hadron block computation) using the helium-3 I/II operators as in Eq. (\ref{eqn:ops}). Figure \ref{fig:tree_benchmark} demonstrates that factor trees offer between one and two orders of magnitude improvement over unoptimised correlator computations for light nuclei using Hadron Blocks. An up-front computational cost for constructing the factor tree not included in Figure \ref{fig:tree_benchmark} is required and proportional to the unoptimised correlator computational cost. The extra calculation, however, is only required before the first configuration is analysed and hence the cost may be amortised over the full set of configurations. Although the factor tree method shows promising speed-ups for small nuclei, it should be noted that the memory used will become prohibitive for large nuclei\footnote{Constructing factor trees directly to hard-disk could extend the method to larger nuclei.}.

\section{Tensor E-graphs} \label{sec:TensorEgraphs}

\textit{E-graphs} (equality graphs) compactly represent and compute a congruence relation over a set of expressions defined by a \textit{re-write rule}. E-graphs were originally developed for automated theorem provers, and have recently shown success when used in program optimisation \cite{willsey_2021}. An \textit{e-graph} is a tuple $(U,M,H)$ where:
\begin{itemize}
    \item $U$ is a union-find data structure (see Refs.\ \cite{willsey_2021,tarjan_1975}) storing an equivalence relation over \textit{e-class ids} (an \textit{e-class} is a set of equivalent \textit{e-nodes}; an \textit{e-node} is a function symbol paired with a list of children e-classes),
    \item $M$ is a map from \textit{e-class ids} to \textit{e-classes}; all equivalent \textit{e-class ids} map to the same \textit{e-class},
    \item $H$ is a hash-cons (see Ref.\ \cite{willsey_2021}) mapping \textit{e-nodes} to \textit{e-class ids}.
\end{itemize}
\par
\begin{figure}[H]
    \centering
    \includegraphics[width=0.5\textwidth]{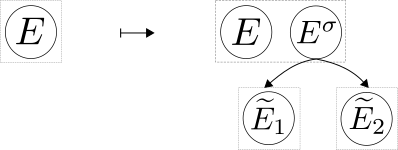}
    % \caption[Formally, $S_{|E|}^\pm$ is the hyperoctohedral group defined by external wreath product $\mathbb{Z}\ wr\ S_{|E|}$]{Tensor e-graph re-write rule for a tensor expression $E$, with $\sigma \in S_{|E|}^\pm$ (signed symmetric group defined by the external wreath product $\mathbb{Z}_2\ wr\ S_{|E|}$ for symmetric group $S_{|E|}$, see Ref.\ \cite{baake_1984}) permuting the index slots of $E$ through action $E^\sigma = E_1 E_2$ for subexpressions $E_1$, $E_2$ related to child e-nodes through tensor expression canonicalisation $E_k \to \widetilde{E}_k$ (see Ref.\ \cite{li_2018}). E-nodes are notated by circles and e-classes are notated by dashed rectangles.}
    \caption{Tensor e-graph re-write rule for a tensor expression $E$, with $\sigma \in S_{|E|}^\pm$ (signed symmetric group\protect\footnotemark) permuting the index slots of $E$ through action $E^\sigma = E_1 E_2$ for subexpressions $E_1$, $E_2$ related to child e-nodes through tensor expression canonicalisation $E_k \to \widetilde{E}_k$ (see Ref. \cite{li_2018}). E-nodes are notated by circles and e-classes are notated by dashed rectangles.}
    \label{fig:egraph_rw1}
\end{figure}

 Here we present \textit{tensor e-graphs}, in which each e-node is a tensor expression $E$ and the initial e-graph is the set of tensor expressions representing the correlator $C$. The full tensor e-graph is built by repeatedly applying to each e-node $E$ the re-write rule given in Figure \ref{fig:egraph_rw1} to produce candidate common subexpressions $\widetilde{E}_1$, $\widetilde{E}_2$.

\footnotetext{Formally, $S_{|E|}^\pm$ is defined by the external wreath product $\mathbb{Z}_2 \wr S_{|E|}$ for symmetric group $S_{|E|}$; see Ref.\ \cite{baake_1984}.}

\noindent In order to restrict the rather large class of possible re-writes $E \xrightarrow{\sigma} E^\sigma = E_1 E_2$, we introduce two conditions:
\begin{enumerate}
    \item Both $E_1$ and $E_2$ must contain at least one summed index
    \item At least one of $E_1$ or $E_2$ must contain at least one fully summed tensor.
\end{enumerate}

\begin{figure}[H]
    \centering
    \includegraphics[width=0.5\textwidth]{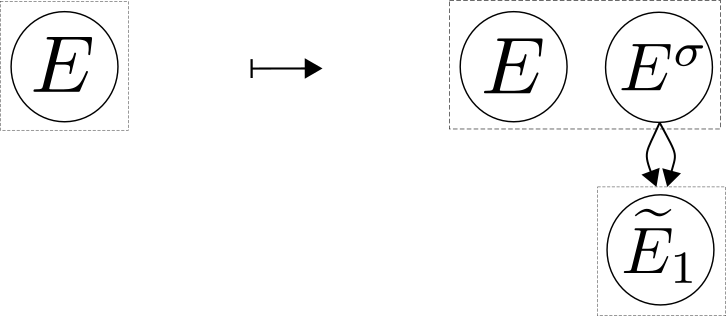}
    \caption{Tensor e-graph re-write rule for tensor expression $E = \epsilon_{ijk} \epsilon_{lmn} T_{ijl} T_{nmk}$ with a rank-3 tensor $T$. Common sub-expression $\widetilde{E}_1 = \epsilon_{ijk} T_{ijl}$ only appears once in the e-graph, enabling its re-use.}
    \label{fig:egraph_rw2}
\end{figure}

For example, take $E = \epsilon_{ijk} \epsilon_{lmn} T_{ijl} T_{nmk}$ for any rank-3 tensor $T$ and consider the sub-expressions $E_1 = \epsilon_{ijk} T_{ijl}$ and $E_2 = \epsilon_{lmn} T_{nmk}$. The canonical forms of these sub-expressions are $\widetilde{E}_1 = \epsilon_{ijk} T_{ijl} = \widetilde{E}_2$, noting that $E_2 \to \widetilde{E}_2$ introduces a relative minus sign from the exchange of anti-symmetric indices $\epsilon_{lmn} = -\epsilon_{nml}$. The signed permutation $\sigma$ keeps track of both the rearrangement of tensors $\epsilon \epsilon T T \to \epsilon T \epsilon T$ and the relative sign induced by the tensor expression canonicalisation process. In this case, $\sigma = -(4\ 7)(5\ 8)(6\ 9)(7\ 9)$ using (signed) cycle notation. This re-write, as depicted in Figure \ref{fig:egraph_rw2} can be used to extract a common sub-expression $B_{kl} = \epsilon_{ijk} T_{ijl}$ so that by permuting the index slots of $E$ by $\sigma$ (i.e. $ijklmnijlnmk \to -ijkijlnmlnmk$) and discarding the index slots summed in $B$, we may express $E = -B_{kl} B_{lk}$.

A tensor e-graph is constructed by repeated application of the re-write rule until the \textit{saturation limit}, where no further non-redundant re-writes may be performed, is reached. Often, it is not computationally feasible to reach the saturation limit, and the tensor e-graph construction process terminates after a set e-node limit has been reached. \par
After a tensor e-graph has been constructed, the optimal sub-expression decomposition scheme (i.e. the desired computational strategy) may be extracted by making a recursive selection within each e-class of the e-node maximising an objective function. Here we use the objective function such that when evaluated on an e-node is given by the number of excess parents from distinct e-classes, multiplied by the number of floating point operations required to evaluate the tensor expression represented by that e-node. This objective function is chosen to maximise the re-use of common sub-expressions weighted by how expensive they are to compute, so as to minimise the total computational cost. The provably optimal objective function remains open. \par

Applying this process to the correlator for the dinucleon II operator as in Eq. (\ref{eqn:ops}) yields the following two example common subexpressions:
\begin{align}
    &\mathcal{B}_1(\alpha', \beta', \gamma', \delta') = \epsilon_{abc} \epsilon_{def} (C\gamma_5)_{\alpha \beta} (C\gamma_5)_{\gamma \delta} f^P_{\alpha, \vec 0}(\vec 0, a, \alpha', d, \beta', b, \gamma') f^N_{\beta, \vec 0}(\vec 0, e, \gamma, c, \delta', f, \delta), \\
    &\mathcal{B}_2(a',b',c',d',e',f') \notag \\
    &\qquad \quad =(C\gamma_5)_{\alpha \beta} (C\gamma_5)_{\gamma \delta} (C\gamma_5)_{\sigma \rho} (C\gamma_5)_{\mu \nu} f^P_{\alpha, \vec 0}(\vec 0, a', \gamma, b', \sigma, c', \rho) f^N_{\beta, \vec 0}(\vec 0, d', \mu, e', \delta, f', \nu).
\end{align}
Note that both $\mathcal{B}_1$ and $\mathcal{B}_2$ satisfy the two re-write restrictions since they both contain  the fully summed tensor $(C\gamma_5)_{\alpha \beta}$, for example. The correlator for the dinucleon II operator is optimally computed by first computing $\mathcal{B}_1$ and $\mathcal{B}_2$, and then contracting the remaining indices. Future work will measure the performance improvement attained by reusing $\mathcal{B}_1$ and $\mathcal{B}_2$ in the contraction.

\section{Conclusion and Outlook}

In this work, we have explored a number of ways to accelerate the computation of correlation functions of multi-hadron systems in the context of lattice QCD. The construction of factor trees shows promising speed-ups for certain choices of interpolating operators of light nuclei once the tree construction cost has been amortised over several iterations, but its applicability to heavier nuclei is constrained by memory limitations. The bulk of the correlator cost for interpolating operators of light nuclei remains in the hadron block evaluation. The construction of optimal evaluation schemes through tensor e-graphs is a potential avenue for performance improvements for larger nuclei, although rigorous demonstration of this possibility is left for future work.

\section*{Acknowledgements}
We would like to thank Artur Avkhadiev and Phiala Shanahan for many helpful discussions. The calculations were carried out on the NCI National Facility in Canberra, Australia (supported by the Australian Commonwealth Government) and the CSSM/HEP HPC cluster (University of Adelaide). NH is supported by an Australian Government Research Training Program (RTP) Scholarship. RDY and JMZ are supported in part by the Australian Research Council grant DP190100297. WD is supported in part by the U.S.~Department of Energy, Office of Science, Office of Nuclear Physics under grant Contract Number DE-SC0011090,  by the SciDAC4 award DE-SC0018121, and by the National Science Foundation under Cooperative Agreement PHY-2019786 (The NSF AI Institute for Artificial Intelligence and Fundamental Interactions, \url{http://iaifi.org/}).

\bibliographystyle{JHEP}
\bibliography{refs}

\end{document}